\documentclass[useAMS,usenatbib,usegraphicx]{mn2e}

\usepackage{amsmath}

\title[Constraints on sterile neutrinos in Segue 1]{{\it Swift} 
observation of  Segue 1: constraints  
on sterile neutrino parameters in
the darkest galaxy}
\author[Mirabal]{N. Mirabal$^{1,2}$\thanks{E-mail:
mirabal@gae.ucm.es}\\
$^{1}$Ram\'on y Cajal Fellow\\
$^{2}$ Dpto. de F\'isica At\'omica,
Molecular y Nuclear, Universidad Complutense de
Madrid, Spain\\}

\begin{document}

\date{}

\pagerange{\pageref{firstpage}--\pageref{lastpage}} \pubyear{2010}

\maketitle

\label{firstpage}

\begin{abstract}
Some extensions of standard particle physics 
postulate that dark matter may be partially
composed of weakly-interacting sterile neutrino particles that have 
so far eluded detection. 
We use a short ($\sim 5$ ks) archival
X-ray observation of Segue 1 obtained with 
the X-Ray Telescope (XRT) on board the {\it Swift} satellite to exclude the 
presence of sterile neutrinos 
in the 1.6--14 keV mass range down to a flux limit of
$6 \times 10^{-12}$ ergs cm$^{-2}$ s$^{-1}$ within 67 pc of its center. 
With an estimated mass-to-light ratio of 
$\sim$ 3400 M$_{\odot}$/L$_{\odot}$, Segue 1 is 
the darkest ultra-faint dwarf galaxy currently measured. Spectral
analysis of the {\it Swift} XRT data fails to find any 
non-instrumental spectral feature possibly connected with the radiative 
decay of a dark matter particle. Accordingly, we establish 
upper bounds on the sterile neutrino
parameter space based on the non-detection of emission lines in the 
spectrum. The 
present work provides the most sensitive X-ray search for 
sterile neutrinos in a region with the 
highest dark matter density yet measured.

\end{abstract}

\begin{keywords}
X-rays: general 
-- galaxies: Local Group -- cosmology: dark matter 
-- galaxies: individual (Segue 1)
\end{keywords}

\section{Introduction}
The nature of dark matter is possibly the greatest
mystery of present-day science. 
Nearly 80\% of the mass density
in the Universe cannot be explained with ordinary baryonic matter and
requires an additional non-baryonic component aptly named dark matter to
reconcile the inferred mass budget 
\citep{gaitskell}.
Barring a radical paradigm shift in fundamental physics,
dark matter is expected to encompass one or more species of
undiscovered elementary particles that would account for the largest
fraction of mass in the Universe. The direct hunt for the
culprits is ongoing at various particle physics experiments in the 
predicted
range for supersymmetric particles at center-of-mass energies 
between 10 GeV and a few TeV \citep{gaitskell}.
Indirect dark matter signals from pair annihilations or
particle decays are also being sought in a similar
range through gamma-ray observations of astrophysical objects with 
high dark matter density \citep{aliu,abdo}. 
However, no significant dark matter
signal has been verified \citep{anderson}.

Given our current blindness to the nature of the dark matter particle, it is 
crucial to span the entire range of hypothesised 
candidates/signatures of dark matter including those that lie 
outside the gamma-ray regime. One
such possibility is a
sterile neutrino, which corresponds to a weakly-interacting
fermion associated with the neutrino sector
that arise in certain extensions of the standard model
\citep{dodelson}. A large accumulation of sterile neutrinos in a 
very compact region could produce
a detectable X-ray flux made primarily of
particle decays
into mono-energetic photons in the 0.1--100 keV energy range
\citep{abazajian,dolgov,yuksel}. Like other dark matter candidates, 
any direct or indirect detection of sterile neutrinos will be a 
challenging but necessary step for our assessment of dark matter in
the Universe \citep{gelmini}.

In order to complete a stringent sterile neutrino search, 
one would like to target astrophysical objects with the highest
dark matter densities. Taken at face value, 
galaxy clusters should 
be excellent places to conduct such searches, but their
cores are filled with hot emitting interstellar medium that would 
veil any X-ray trace of
a dark matter particle. Similar issues arise in the X-ray emission from 
massive galaxies that tends to be dominated by hot
gas and binary systems. The next best targets may be
hiding among
the recently discovered population of ultra-faint 
dwarf galaxies around the Milky Way \citep{willman1,belokurov}.
Progress is still being made in characterising the  
kinematic properties
of these objects \citep{simon}. However, spectroscopic 
observations suggest that some ultra-faint dwarf galaxies could be the
closest and densest dark matter haloes in the Local
Group, making them excellent targets to
conduct searches for the annihilation or decay of dark matter
particles at high energies \citep{strigari,simon}.

Indirect searches for sterile neutrinos in dwarf galaxies have been
conducted in Fornax \citep{boyarsky}, Ursa Minor \citep{loew2},
and Willman 1 \citep{loew,mirabal}. Out of this group, only Willman 1 
falls under the
category of ultra-faint dwarf galaxy ($L < 1000L_{\odot}$) 
and it is one of the 
preferred targets for dark matter pursuits \citep{strigari}. 
However, careful interpretation of Willman 1 has found
strong evidence for tidal disruption 
 and possible contamination from foreground stars that could 
prevent any reliable determination of its dark matter mass from stellar 
kinematics 
 \citep{siegel,willman2}.

In order to sidestep the kinematic complications surrounding Willman 1, 
we turn to the Milky Way satellite Segue 1 \citep{belokurov}. 
At a distance of 23 kpc, 
Segue 1 is the closest ultra-faint dwarf galaxy. With the highest
mass-to-light ratio  ($\sim$ 3400 M$_{\odot}$/L$_{\odot}$)
and dark matter density $2.5^{+4.1}_{-1.9}$ M$_{\odot}$ pc$^{-3}$
yet observed, there is general consensus that 
Segue 1 is the next most promising object for a potential indirect 
detection of dark matter
\citep{strigari,simon}. 
In this Letter, we use a {\it Swift} X-ray observation to
constrain the parameter space of sterile neutrinos in Segue 1.

\begin{figure}
\hfil
\includegraphics[width=3.1in,angle=0.]{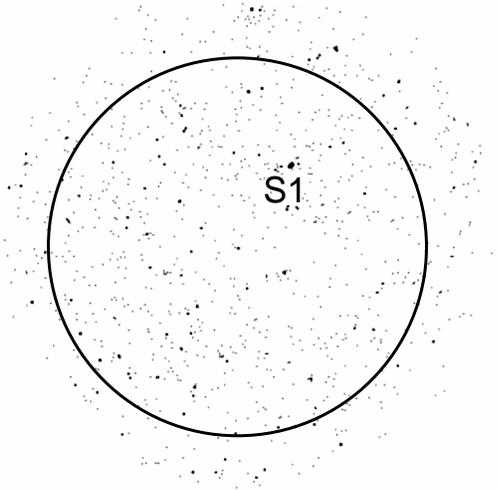}
\hfil
\caption{Smoothed {\it Swift} XRT 0.2--10 keV image of Segue 1. The circle
represents the circular extraction region of radius 10\arcmin\  (67 pc) 
centered
on (J2000.0) RA. $10^{\rm h}07^{\rm m}03.\!^{\rm s}2$,
Dec. $+16^{\circ}04\arcmin25\arcsec$. S1 marks the position of
SWIFT J100651.4+160845 the single point
source in the field detected with signal-to-noise ratio
$> 5$.  
}
\label{figure1}
\end{figure}

\section{Observation and X-ray source detection}
A continuous 4,855 s exposure 
of Segue 1 was acquired on 
2010 February 7 with the X-ray telescope (XRT; Gehrels et al. 2004)
on-board the {\it Swift}
satellite. 
The XRT observation (ID 00031602003) was 
collected in  photon counting (PC) mode with full spectral resolution 
and a time resolution of 2.5 s. For the analysis,
we used the cleaned event file with standard grade filtering (grades 0--12) 
and default screening
parameters in the 0.2--10 keV energy range resulting in an
effective exposure time of 4,834 s.  Analysis of the data
was performed with standard HEAsoft and CIAO tools. Fig. \ref{figure1}
shows the XRT image of the field. The image has been smoothed
with a Gaussian kernel $r_{k}$ = 7\arcsec. At 
Galactic coordinates $\ell = 220.5^\circ$, $b= 50.4^\circ$,
Segue 1 lies well away from 
the Galactic plane with Galactic H I
column density $N_{\rm H}$ =
$3.3 \times 10^{20}$ cm$^{-2}$ \citep{dickey} and 
minimal contamination expected 
from foreground Galactic point sources.

A thorough census of point sources in the field was performed using 
the Mexican-hat wavelet routine $wavdetect$ on scales of 6, 10, and
16 pixels \citep{freeman}. Detections of X-ray point sources required 
a signal-to-noise ratio
$> 5$. A single 
source SWIFT J100651.4+160845 (S1) meeting our requirement was found  
at (J2000) $10^{\rm h}06^{\rm m}51.\!^{\rm s}4$,
Dec. $+16^{\circ}08\arcmin45.3\arcsec$ with an uncertainty of
5\arcsec. Within a 20 pixel radius centered on SWIFT J100651.4+160845
the 0.3--7 keV count rate is $(4.1 \pm 0.9) \times 10^{-3}$ s$^{-1}$
corresponding to an extrapolated flux  in the 0.2--10 keV band
of $(1.5 \pm 0.3) \times 10^{-13}$ ergs cm$^{-2}$ s$^{-1}$
with photon index $\Gamma = 2.0$ and $N_{\rm H}$ =
$3.3 \times 10^{20}$ cm$^{-2}$ as obtained from WebPIMMS
\footnote{http://heasarc.gsfc.nasa.gov/Tools/w3pimms.html}. 

At the derived X-ray position of 
SWIFT J100651.4+160845, we find SDSS J100651.56+160847.6  
a magnitude of $r = 20.1$ optical object
that has been photometrically typed as a potential quasi-stellar object
(QSO) by the Sloan Digital Sky Survey (SDSS) pipeline \citep{abazajian2}.
The X-ray point-source catalogue from the {\it Chandra} Deep Field-South 
(CDF-S) also predicts at least one background QSO unrelated to
Segue 1 to this sensitivity limit 
for a {\it Swift} area of $\approx 314~arcmin^{2}$ \citep{luo}. 
Accordingly, we assign a tentative background QSO classification to 
SWIFT J100651.4+160845.

\section{Spectral analysis and sterile neutrino parameters}
Before extracting the actual spectrum of Segue 1, we excised
SWIFT J100651.4+160845 from the event list. 
We obtained the spectrum using a  
10\arcmin circle radius centered on
the position of Segue 1 (J2000.0) RA. $10^{\rm h}07^{\rm m}03.\!^{\rm s}2$,
Dec. $+16^{\circ}04\arcmin25\arcsec$
determined by \citet{martin}. The radius corresponds to  
a physical size of 67 pc at the distance of 23 kpc, equivalent to
2.3 half-light radii
\citep{simon}.
The resulting spectrum contains 650 net counts in the 0.2--7 keV band. 
Since the extraction region nearly covers the totality of
the {\it Swift} XRT field, we obviate any background subtraction. 
We reason that for the desired upper bound on the 
parameter space of sterile neutrinos of Segue 1, it is better to provide
a conservative estimate that includes 
the background contribution rather than to gamble with an
inadequate instrumental background subtraction 
\citep[see a similar argument by][]{riemer}.
It is important to note that 
an extrapolation of the counts in the area 
outside the extraction region indicates
that the purported Segue 1 emission has signal-to-noise ratio smaller
than 3.

\begin{figure}
\hfil
\includegraphics[width=3.1in,angle=0.]{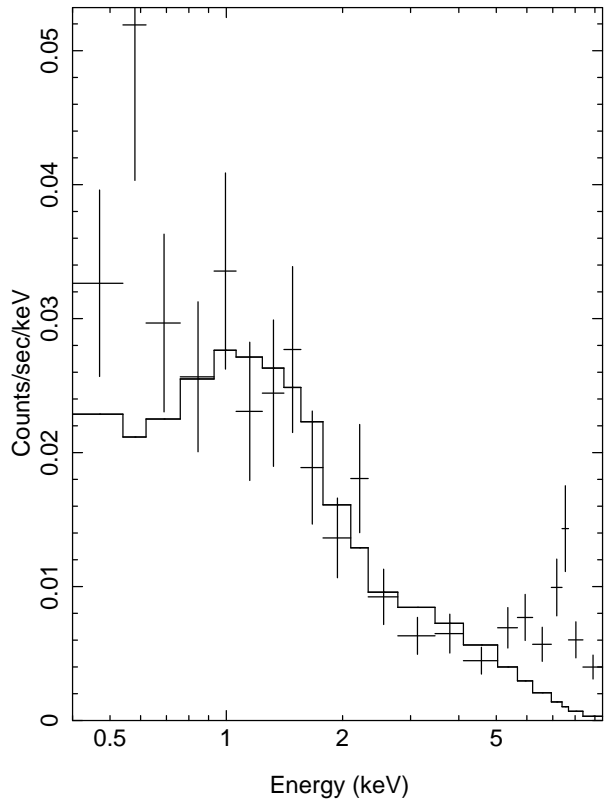}
\hfil
\caption{
{\it Swift} XRT 
spectrum of Segue 1 in the 0.4--10 keV range. The spectral fit correspond to
an absorbed power law $\Gamma = 1.4 \pm 0.3$ with Galactic H I
column density $N_{\rm H}$ =
$3.3 \times 10^{20}$ cm$^{-2}$. 
Excesses at 0.6, 2.3 and 7 keV are likely due to instrumental oxygen,
gold, and nickel edges respectively. 
}
\label{figure2}
\end{figure}

The extracted spectrum was grouped with $grppha$
to ensure at least 20 counts per bin. Subsequently, an 
ancillary response function (ARF) was generated through $xrtmkarf$ using 
the response matrix file (RMF) from 
the latest distribution in {\it Swift} Calibration 
Database\footnote{http://heasarc.gsfc.nasa.gov/docs/heasarc/caldb/swift/}.
We then fitted the spectrum with a simple absorbed power-law model 
implemented in XSPEC \citep{arnaud}.
Between 0.4 and 5 keV, the spectrum is reasonably fit (reduced 
$\chi^{2} = 1.04$)
with a power-law index 
$\Gamma = 1.4 \pm 0.3$ with N$_{\rm H}$ fixed at the Galactic value. 
Fig. \ref{figure2} shows the compounded (source + instrumental background)
spectrum of Segue 1 with a power law.

As shown in this Fig. \ref{figure2}, we find no
evidence for the presence of emission lines  
in the compounded spectrum of Segue 1.
The main residuals at 0.6 and 2.3 keV between the spectrum and power-law model
have been identified with instrumental 
oxygen and gold edges respectively \citep{godet}. Nickel contamination
is also likely present in the 7--8 keV range \citep{cusumano}. 
For dark matter applications, we find no conclusive 
evidence of excess emission at $\sim 
2.5$ keV as previously reported
in the spectrum of Willman 1 \citep{loew}. However, we must point out that 
this shorter {\it Swift} XRT observation was less sensitive and cannot
completely rule out a line with flux properties $F_{\rm line} 
= 3.5 \times 10^{-6}$ ${\rm ph}~{\rm cm}^{-2}~{\rm s}\
^{-1}$ claimed by
those authors.

\begin{figure}
\hfil
\includegraphics[width=3.1in,angle=0.]{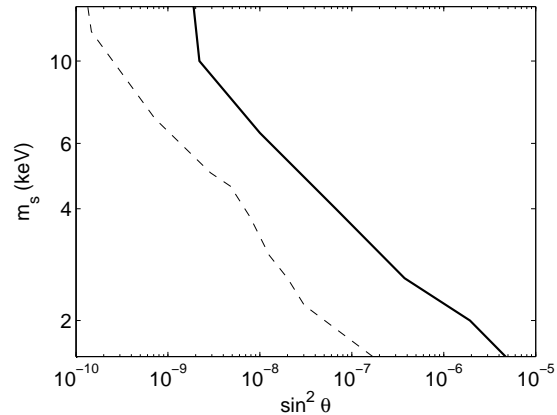}
\hfil
\caption{Sterile neutrino parameter space. The thick solid line represents
the constraints on sterile neutrinos derived
for Segue 1. The thin dashed line corresponds to the upper bound 
in Willman 1. The parameter space to the right of each line
marks the exclusion region assuming
100\% of dark matter is composed of sterile neutrinos.  
}
\label{figure3}
\end{figure}

In the absence of actual spectral features, the next best thing to do
is to produce upper bounds for the sterile neutrino parameter space
that directly depend on the upper limits to X-ray counts from 
emission lines $F_{\rm line}$. For this purpose, 
we fitted Gaussian lines fixed at a line width $\sigma = 100$ eV 
between 0.8 and 7 keV using XSPEC. Line emission upper limits were calculated 
in steps of 0.5 keV over said energy range. 
We adopt a conservative posture and assume that 
only line emission (no underlying continuum or instrumental background) 
contributes to the observed
line flux at each energy. In order to convert
these measured emission line upper limits into actual points in the 
sterile neutrino parameter space (mass m$_{s}$ 
versus mixing angle $\theta$), we adopt the formalism
of \citet{loew},

\begin{equation}
F_{\rm line} = 5.15~sin^{2}\theta~
\left(\frac{m_{s}}{\rm keV}\right)^{4}
\times f_{s} {\rm M}_{7} {\rm d}^{-2}_{100}~{\rm ph}~{\rm cm}^{-2}~{\rm s}^{-1}
\end{equation}

where sterile neutrinos with mass $m_{s}$  produce 
photons at a given line energy $E_{\rm line} = m_{s}/2$,
M$_{7}$ is the dark matter mass of Segue 1 
in units of $10^{7}$ M$_{\odot}$, $f_{s}$
is the fraction of dark matter in sterile neutrinos, and d$_{100}$ is the 
distance to Segue 1 in units of 100 kpc. 
For the actual calculations, we assume that 100\% of
dark matter in Segue 1 is composed of sterile neutrinos $f_{s} = 1$,
a helio-centric distance d$_{100} = 0.23$ \citep{simon}, 
and a dark matter mass M$_{7} = 0.06$ \citep{wolf,simon}. Fig. \ref{figure3}
shows the derived sterile neutrino mass m$_{s}$
as a function of mixing angle $\theta$. Also shown 
are the corresponding upper bounds obtained for Willman 1 \citep{loew,mirabal}.
It should be noted that there is considerable uncertainty on the
the dynamic modelling of Willman 1 \citep{willman2}.

The constraints for Segue 1 shown in Fig. \ref{figure3} are about a factor of
10 less sensitive 
than the upper limits derived for Ursa Minor \citep{boyarsky2}.
Furthermore, the results are even less restrictive than those reported for
the Milky Way \citep{boyarsky2}, M31 \citep{watson}, and the unresolved
cosmic X-ray background \citep{aba07}. However, to a large extent, 
estimates based on Equation (1) depend exclusively 
on the total dark matter mass 
M$_{7}$ rather than specific dark
matter densities within the object, which may be more appropriate. At 
$2.5^{+4.1}_{-1.9}$ M$_{\odot}$ pc$^{-3}$, the dark matter density of
Segue 1 is significantly (a factor between 2 and 200) 
higher than that measured in any bound system
\citep{simon2,weber}. Assuming that the sterile neutrino line 
emission has an additional
dependence on the dark matter density $n_{\rm dm}$
on small scales, density weighted
upper limits could be of similar magnitude.

\section{Discussion and Conclusions}
Based on the observational limits reported here, 
an unequivocal dark matter signal 
continues to evade us in the X-ray regime. None the less, 
by focusing on regions with high dark matter density,
we appear to
be  on the right path to discovery (or exclusion).
Because no background subtraction was performed on the compounded spectrum
of Segue 1,
the exclusion region for sterile neutrinos 
obtained from this observation 
is less restrictive than previous measurements for similar systems 
\citep{loew,mirabal,boyarsky}. Yet, the exercise is
valuable as the hunt for an indirect dark matter
signal continues from X-ray to gamma-ray energies. It is unfortunate that
the observation cannot confirm the presence of the spectral feature 
claimed in the Willman 1 spectrum because of poorer sensitivity
\citep{loew}. However, the reality of this
marginal spectral feature remains hotly
contested \citep{boyarsky,mirabal}. An additional concern with Willman 1 
is the fact that its kinematic properties make it
difficult to determine its actual dark matter mass and 
carry out any subsequent sterile neutrino calculations
\citep{willman2}. 
%Taken together, these latest results significantly 
%challenge the low significance report of
%a spectral feature connected with sterile neutrinos by \citet{loew}. 
%Therefore, Segue 1 appears to 
%take over as the only standing ultra-faint dwarf galaxy with a
%robust dark matter density determination derived from kinematics.  

From the point of view of kinematics, Segue 1 has also faced  
a serious challenge by \citet{niederste} who
argued that the dark matter
properties of Segue 1 might be inflated due to possible
contamination from stars in the Sagittarius stream.
Under the latter interpretation, unrelated field stars
would disturb the velocity dispersion measurements and hence inflate
the inferred mass-to-light ratio of an otherwise normal stellar cluster.
However, recent work counters that 
with additional stars such possibility  
is nearly ruled out
\citep{simon}. As a result, Segue 1 appears to
qualify as the densest dark-matter dominated object with a
robust determination of sterile neutrino parameters in the 0.8--7 keV 
bandpass.

If the high dark matter density value 
is confirmed with further measurements, Segue 1 will
certainly transform itself in the ideal astrophysical laboratory to 
contrast any direct
dark matter particle detection  at ground-based accelerators. 
But any such effort will require that the
field of view of Segue 1 is certified as ``clean'' 
{\rm i.e.} void of X-ray/gamma-ray contaminants 
that might entangle any interpretation of the
high-energy emission other than obvious background gamma-ray quiet
QSOs. With the discovery of  
SWIFT J100651.4+160845, we have just begun to probe the X-ray
point-source population in the field of Segue 1. 
Given the potential of novel discoveries,
it seems worthwhile to observe Segue 1 with more sensitive
X-ray instruments on board the {\it Chandra} and {\it XMM-Newton} 
observatories that could
conclusively quantify the contamination from
unrelated sources.

\section*{Acknowledgments}
N.M. gratefully acknowledges support from the Spanish Ministry of Science
and Innovation through a Ram\'on y Cajal fellowship. Support from the 
Consolider-Ingenio 2010 Programme under grant MULTIDARK 
CSD2009-00064 is also acknowledged. We thank the referee for comments. 
Finally, we acknowledge the use of public data from the {\it Swift} 
data archive.

\label{lastpage}
\end{document}